# Malaria Parasitic Detection using a New Deep Boosted and Ensemble Learning Framework

Saddam Hussain Khan[1*], Tahani Jaser Alahmadi[2]

[1]Department of Computer Systems Engineering, University of Engineering and Applied Science, Swat, Pakistan
[2]Department of Information Systems, Princess Nourah bint Abdulrahman University, Riyadh, Saudi Arabia

## Abstract

Malaria is a potentially fatal plasmodium parasite injected by female anopheles mosquitoes that infect red blood cells and cause millions of lifelong disability worldwide yearly. However, specialists' manual screening in clinical practice is laborious and prone to error. Therefore, a novel Deep Boosted and Ensemble Learning (DBEL) framework, comprising the stacking of new Boosted-BR-STM convolutional neural networks (CNN) and the ensemble ML classifiers, is developed to screen malaria parasite images. The proposed Boosted-BR-STM is based on a new dilated-convolutional block-based Split Transform Merge (STM) and feature-map Squeezing-Boosting (SB) ideas. Moreover, the new STM block uses regional and boundary operations to learn the malaria parasite's homogeneity, heterogeneity, and boundary with patterns. Furthermore, the diverse boosted channels are attained by employing Transfer Learning-based new feature-map SB in STM blocks at the abstract, medium, and conclusion levels to learn minute intensity and texture variation of the parasitic pattern. Additionally, to enhance the learning capacity of Boosted-BR-STM and foster a more diverse representation of features, boosting at the final stage is achieved through TL by utilizing multipath residual learning. The proposed DBEL framework implicates the stacking of prominent and diverse boosted channels and provides the generated discriminative features of the developed Boosted-BR-STM to the ensemble of ML classifiers. The proposed framework improves the discrimination ability and generalization of ensemble learning. Moreover, the deep feature spaces of the developed Boosted-BR-STM and customized CNNs are fed into ML classifiers for comparative analysis. The proposed DBEL framework outperforms the existing techniques on the NIH malaria dataset that are enhanced using discrete wavelet transform to enrich feature space. The proposed DBEL framework achieved Accuracy (98.50%), Sensitivity (0.9920), F-score (0.9850), and AUC (0.9960), which suggest it to be utilized for malaria parasite screening.

1. **Introduction**

Malaria is a life-threatening illness transmitted by female Anopheles mosquitoes that injects plasmodium parasites with one nasty bite. In most cases, plasmodium parasites target healthy red blood cells (RBC) around one or two weeks after their emergence in the human body [1]. This bacterial infection is hazardous to kids, persons with impaired immune systems, pregnant women, and the elderly at risk [2]. P. Falciparum malaria is particularly dangerous to pregnant women as it increases stillbirth, maternal death, miscarriage, and newborn [3]. The World Health Organization reported roughly 241-million malaria suspects and 627000 fatalities in 2021. The African continent is perhaps the most afflicted, accounting for 95% of 90% of deaths and 80% of child disability caused by acute malaria [4,5].

RBCs were microscopically examined in a thick, thin blood smear frequently used to identify malaria [6]. The thick and thin-smear test aids in identifying the density of parasites in a person's body and malaria species, respectively [7,8]. Expert pathologists manually analyze blood smear films to get a microscopic diagnosis which is time-consuming, laborious, and unreliable [9]. Malaria patients are usually seen in emerging countries, where diagnostic lab facilities and tools are unavailable. In addition, a global shortage of trained professionals significantly impacts the healthcare systems of developing countries [10]. Therefore, a computer-based screening tool is essential for speedy and reliable malaria analysis [11].

Artificial intelligence (AI) and machine learning (ML) have aided in the development of malaria diagnostic methods that are effective and precise in processing large amounts of parasite-contaminated RBC samples [12–14]. Computer-based software will support clinicians in diagnosis and therapy, facilitating established lab practices [15–17]. However, this convention performed poorly on massive data and was incapable of learning complex patterns. Therefore, Deep Learning (DL) algorithms emerged and inspired researchers' interest in dealing with enormous amounts of data and learning complicated patterns [18,19]. They have significant growth for medical imaging infection diagnosis [20–22]. Malaria parasite analysis is crucial for diagnosing infected cells. In this regard, a deep CNN-based identification helps quickly and accurately analyze malaria parasite images. Several CNN-based classification frameworks and experimental models have extensively been employed on the NIH malaria dataset to improve detection [23,24]. CNN extracts deep features automatically by avoiding the time-consuming hand-crafted feature extraction and reduces computational power [23,25].

To our knowledge, this study is the first to introduce a new deep hybrid framework comprised of novel Deep residual and spatial blocks CNN and ensemble learning to analyze malaria parasite-afflicted patients accurately and efficiently. The proposed deep Boosted-BR-STM exploits the channel Squeezing and Boosting (SB) technique with a novel Split Transform Merge (STM) block. Moreover, the developed

STM block of deep Boosted-BR-STM uses the concept of homogenous and heterogeneous. The significant contributions are as follows:

1. A new Deep Boosted and Ensemble Learning (DBEL) framework is proposed comprising a new residual learning based Boosted-BR-STM CNN and ensemble learning for detecting RBCs infected with the plasmodium falciparum using blood smear images. The dataset is initially enhanced and reduced in dimension using a discrete wavelet transform (DWT) to improve computational complexity.

2. The proposed deep Boosted-BR-STM exploits the channel SB technique with a novel STM block. Moreover, the developed STM block uses the concept of homogenous and heterogeneous.

3. The innovative SB concept is carefully integrated into the new STM block at abstract, medium, and conclusion levels effectively capturing the diverse pattern of homogeneous, heterogeneous, contrast, and textural variations of the parasitic cell. SB notion is utilized by merging reduced prominent channels with TL-based extracted additional feature maps to improve Boosted-BR-STM performances. In addition, to improve the learning capacity of Boosted-BR-STM and promote a more diverse representation of features, residual learning-based feature map boosting is achieved at the final stage through TL.

4. The proposed framework grants the inherent properties of diverse prominent and boosted channels to the discriminative feature level and fed to the ensemble of ML classifiers, improving the capability of discrimination and generalization of ensemble learning. The ML classifiers' ensemble effectively reduces feature dimension and improves diverse decision space. Ultimately, early detection reduces the possibility of permanent disabilities.

5. The proposed hybrid DBEL framework performance is compared to existing techniques utilized in NIH-malaria original and enhanced datasets.

The rest of the manuscript is arranged in the following ways. In sections 2 and 3, the remainder of the article discusses related work and the proposed malaria detection framework. Materials and implementation details are provided in Section 4. The performance analysis and outcomes are presented in Section 5. Finally, the article's conclusion is in Section 6.

## 2. Related Work

Several ML techniques were automatically determined in stained RBC samples employing methods over time [26]. ML algorithms such as Extreme-Learning Machine methods, Global White Balance, Logistic Regression, SVM, Adaptive Non-linear Grayscale, and KNN have enhanced classification performance [27,28]. The malaria parasite-infected features like RBCs, color, size, form, and statistical data were extracted and provided to ML algorithms [29,30]. Moreover, the deep belief network technique was used to classify the parasites, which yielded a 90.21% accuracy [31]. The approaches mentioned above, on the other hand, indicated prediction accuracy in the range of 84–94%. Several researchers used deep CNN-based detection that automatically and quickly identifies cells infected with the malaria parasite, achieving a 91.50% detection rate [31-33]. Medical datasets with labels are generally limited in size and thus unsuitable for practical application. Therefore, TL has been utilized in a limited labeled malaria dataset to forecast malaria parasites in RBC [32]. TL-based existing CNNs have been employed to improve and have attained accuracies of 75% to 94% [33–35]. Furthermore, R-CNN for parasite identification has been employed with considerable performance [36].

Previous studies employed VGG16 to identify malaria using the common NIH dataset and achieved 95.96% accuracy [37]. The balanced NIH dataset included 27,556 images and was resized to 224 by 224 pixels. A sequential-tailored CNN was obtained with a 95.90% F1 score, 94.7% sensitivity, and 92.70% accuracy [25]. The existing ResNet50 yielded a 95.40% accuracy using the malaria dataset [38]. Moreover, the custom-made CNN model demonstrated 96.82% accuracy, 96.33% using five convolutional and pooling layers, and 96.82% F1-score [23]. Eventually, a hybrid platform that systematically reduces structural and empirical risk has been recently adopted and achieved a sensitivity of 93.44% and an accuracy of 93.13% [25,32]. In another study, a modified Capsule Network (CapsNet) hybrid screening algorithm has been reported for automated identification and pixel label based classification (segmentation) of malaria parasite-infected RBCs, achieving an accuracy of 98.70% [39]. Moreover, modified YOLOV3 and YOLOV4 models have been employed on a publicly available malaria dataset and achieved an accuracy of 95.46% and 96.14%, respectively [40]. A recent study, employed YOLOv5 and DarKnet-53 Plasmodium Falciparum life stages detection and achieved an accuracy 95.20% and 96.02%, respectively [41]. However, certain limitations to the prior work are: Previous CNN models have been used to detect parasite images designed specifically for natural samples. In contrast, the medical images have distinct patterns and textures of the parasite malaria infection, limiting the model's performance. The prior research primarily focused on the accuracy of the validation dataset. Mostly, the previous studies are evaluated using accuracy; however, the results of a single-measure evaluation of the detecting system could be more reliable. Therefore, the detection rate and F-score are essential indicators of parasite detection. Moreover, testing evaluation measures on a vast, previously unseen test dataset is necessary for determining the detection model's robustness. Furthermore, several models have been evaluated on a limited dataset of images of malaria parasites.

## 3. Malaria Parasite Detection Scheme

The proposed DBEL framework and subsequent prompt effective treatment in sufficient time can detect permanent disabilities from malaria parasite. The microscope analyzes malaria parasites from blood-films stained with various chemical stains [42]. Various scanners, chemical stains, concentrations, and methods can cause color differences. Therefore, analyzing parasite and artifact patterns become challenging and occur substantial margins error. These challenges necessitate designing a new deep CNN and ensemble learning framework to capture complicated patterns and identify malaria infectious microscopic thin smear samples. The developed framework comprised a new Boosted-BR-STM, ensemble learning, and contemporary customized CNNs to screen parasite-infected cells from healthy images. The proposed malaria detection scheme utilizes three experimental arrangements: (1) the proposed DBEL and (2) the DBML framework, and (3) the evaluation of existing CNNs. These existing CNNs learned from scratch and fine-tuned using TL on the malaria NIH dataset. Moreover, these CNNs are evaluated as (i) The deep features are extracted from existing CNNs and provided to the ML classifier (ii) The Softmax-based evaluations. Additionally, the DWT and data augmentation as pre-processing techniques for improving the detection CNNs performance to screen stain impurities or artifacts from parasite images and analyze different parasitic patterns. The overall workflow is arranged in Figure 1.

### 3.1. Data Enhancement

Discrete wavelet transformations (DWT) effectively produce salient feature maps and reduce noise [43]. DWT divides the image into four sub-parts: high, low, and diagonal-resolution channels. The next level of DWT, the low and diagonal-resolution channels, is stored and altered to provide the necessary DWT coefficients for parasite detection. Moreover, a 75%-dimension reduction at each level minimizes computational complexities. The low- and high-resolution images are combined and reconstructed using IDWT to produce highly informative (low-resolution and diagonal features) enhanced images [44]. The original and improved DWT data is shown in Figure 2.

### 3.2. Data Augmentation

Data augmentation is widely applied to boost the robustness of the model and lessens the risk of overfitting. Data instances are increased in runtime (on-the-fly augmentation) during the model training using different data augmentation techniques [45]. This work uses rotation, reflection, shearing, as shown in Table 1. Moreover, scaling and translation parameters are kept by default [1 1] and translation [±5], respectively. Data augmentation strategies are employed to learn different variations and improve the model's generalization, especially, shearing learn distortions, real-world variations, and occlusions.

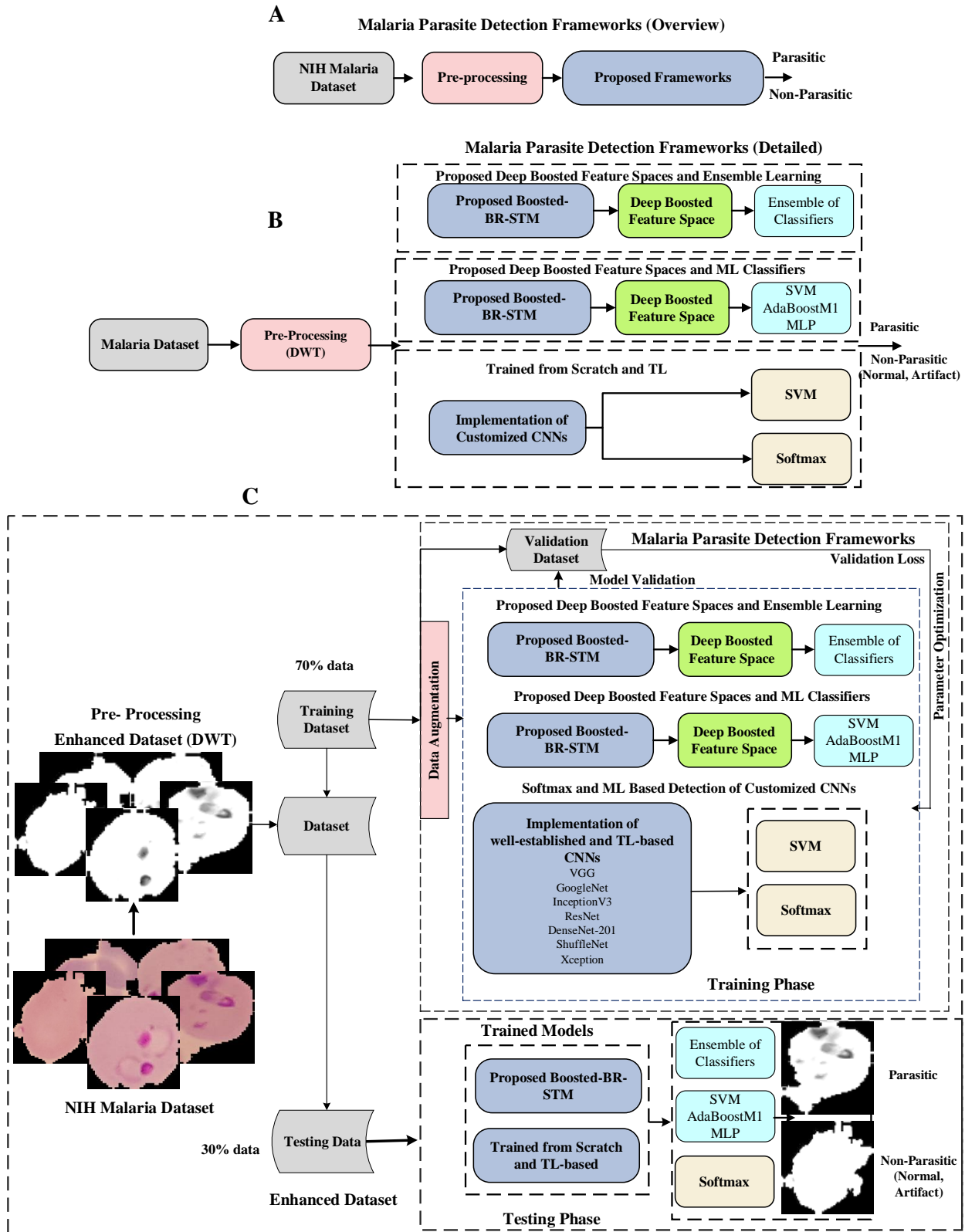

**Figure 1** The flow diagram of the developed malaria parasite detection scheme

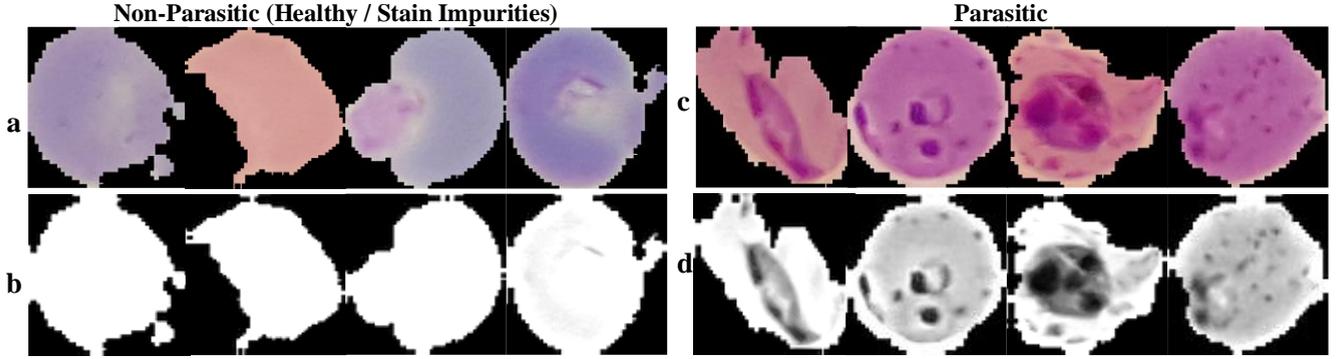

**Figure 2** Parasite and Non-Parasite (healthy, stain artifact) samples from the NIH malaria dataset are shown in panels (a, c). While (b, d) displays their DWT-enhanced examples, respectively.

**Table 1** Data Augmentation detail.

| Parameter | Values |
|---|---|
| Rotation | [± 30] degree |
| Shearing | [0 30] |
| X-Y Reflection | [±1] |

### 3.3. The Proposed Deep Boosted and Ensemble Learning Framework

This research proposes a new deep DBEL framework that is comprised of the developed Boosted-BR-STM BRNet and ensemble techniques for malaria parasite detection. Moreover, the developed Boosted-BR-STM penultimate layers are used for deep feature extraction. The workflow of malaria parasite detection framework is illustrated in Figure 3.

### 3.3.1. The Proposed Channel Boosted-BR-STM CNN

This study introduces a novel Boosted-BR-STM CNN for the detection of parasitic malaria in patients. The CNN incorporates a unique STM-based convolution block to analyze both homogenous and boundary patterns in parasitic regions. Additionally, we employ novel SB concepts within STM to produce condensed and diverse feature maps. The channel SB technique is applied at the initial, intermediate, and final stages to capture textural differences between parasites and typical artifacts [46], as illustrated in Figure 3.

The Boosted-BR-STM effectively learns abstract and target features by organizing three STM blocks with a consistent structure. Each block consists of four dilated-convolutional units, employing max and average pooling to explore border determination, regional consistency, and textural variation [47,48], as shown in the equations (1-3). This systematic implementation aids in efficiently exploring border or edge determination, regional homogenous, and textural variation of parasitic and artifact images.

$$x_{k,l} = \sum_{i=1}^{m} \sum_{j=1}^{n} x_{k+i-1, l+n-1}\, f_{i,j} \qquad (1)$$

$$x^{max}_{k,l} = max_{i=1,\ldots,w,\, j=1,\ldots,w}\, x_{k+i-1, l+j-1} \qquad (2)$$

$$x^{avg}_{k,l} = \frac{1}{w^2} \sum_{i=1}^{w} \sum_{j=1}^{w} x_{k+i-1, l+j-1} \qquad (3)$$

In the equation, '**x**' represents the input feature map, size is shown with the symbols 'k' x 'l' and 'I' x 'j', and 'f' stands for the filter (1) in equations (2-3). As indicated in the equation, every one of the four convolutional blocks (B, C, D, and E) uses channel SB differently to learn distinct parasitic feature sets (4). While learning blocks D and E from scratch, TL builds additional channels in blocks B, C, M, and N to provide different feature maps. Each STM convolutional block has 32, 64, 128, and 128, 256, 512 channel dimensions when squeezed and boosted, correspondingly [49]. Finally, a systematic approach has been adopted, which involves the stacking of TL-based residual learning-based M and N blocks, culminating in their concatenation at the final stage to effectively learn diverse feature spaces. Three residual blocks are sequentially arranged to facilitate the acquisition of diverse features, wherein the number of channels progressively escalates from 32 to 256.

TL's core role is to gain information from the trained source-domain and solve issues in the target-domain while pursuing a high level of performance. Block A also uses region smoothing techniques to minimize the distortion and outlier acquired while capturing the input images [50]. The boosted channel is handled in block F to lessen connection intensity and obtain ideal attributes.

$$\mathbf{x}_{Boosted} = b(\mathbf{x}_B || \mathbf{x}_C || \mathbf{x}_D || \mathbf{x}_E) \qquad (4)$$

$$\mathbf{x}_{DBF} = \sum_{a}^{A} \sum_{b}^{B} v_a \, \mathbf{x}_{Boosted} \qquad (5)$$

$$\sigma(\mathbf{x}) = \frac{e^{x_i}}{\sum_{i=1}^{c} e^{x_c}} \qquad (6)$$

The feature-maps of block D and E are denoted in equation (4) by the variables $x_D$ and $x_E$, respectively. Likewise, blocks B and C is based on additional channels created using TL and are shown as $x_B$ and $x_C$, respectively. $b(.)$ represents the boosting operation $b(.)$. Additionally, the developed CNN employs fully connected layers with dropout layers to collect and preserve target-specific features and minimize overfitting. $v_a$ serve as an example of the number of neurons in equation (5). Lastly, equation (6), whereas c stands for the number of classes, represents softmax, an activation function.

**Significance of using Auxiliary Channels and Squeezing-Boosting (SB) Ideas**

The hybrid framework's representational capacity is enriched by introducing multiple additional channels into the advanced deep Boosted-BR-STM through ensemble learning. The Squeezing-Boosting (SB) ideas employed in the proposed deep CNNs, initially squeezed channels to get salient and informative feature space. Then, combined with each STM block at abstract, mid, and high levels to boost and achieve diverse feature map using TL-based auxiliary channels. Concatenating the prominent and notable information from various deep CNNs using distinct channels enhances the malaria infection depiction.

Moreover, feature-map enhancement is achieved by incorporating auxiliary channels from pre-trained residual models. These supplementary channels are subsequently employed in conjunction with the proposed CNN at the final stage. The SB-based deep CNN effectively learns intricate local and global patterns, enabling the discrimination of textural variations between parasitic and healthy samples.

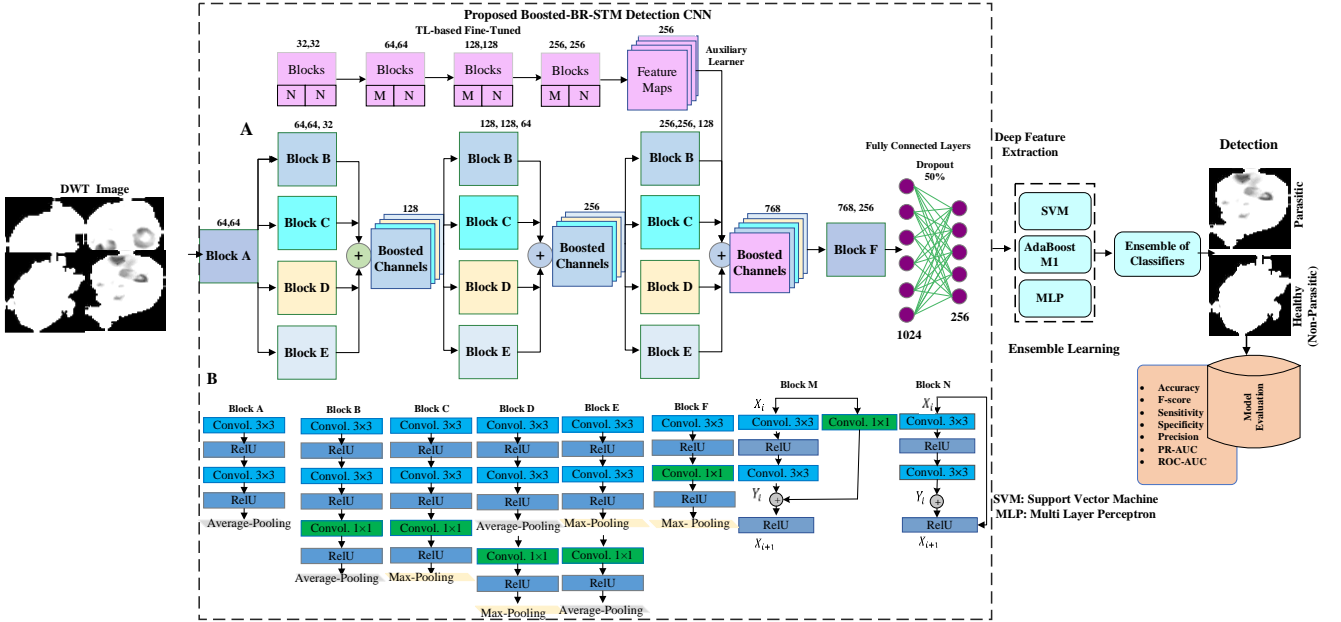

**Figure 3** The proposed malaria parasite detection framework comprised the developed Boosted-BR-STM and Ensemble of Classifiers.

### 3.3.2. Ensemble Learning

We employ deep boosted feature-maps and classifier ensemble strategies in the proposed hybrid learning. The essence of this boosted hybrid learning lies in enhancing the feature space used for combining ML classifiers. Ensemble classifiers, guided by a voting strategy, amalgamate decisions from various classifiers. This hybrid learning strategy harnesses the strengths of the developed deep Boosted-BR-STM and the discriminating power of ML classifiers by isolating key characteristics. We extract deep feature spaces from the final layers of boosted deep CNNs and feed them to competitive ML classifiers, including SVM [48], MLP [51], and AdaBoostM1 [52]. The activation functions are represented in $f_{SVM}(.)$, $f_{MLP}(.)$, $f_{Adaboost}(.)$ as shown in equation (7-10).

The proposed DBEL framework extracts deep boosted channels from the proposed Boosted-BR-STM to achieve diverse channels and provides to ensemble classifier. The ensemble method combines the outputs of individual models, leveraging their diverse strengths. The ensemble aggregates predictions from multiple models, reducing the risk of overfitting and enhancing the generalization ability of the proposed framework. Moreover, the DBEL benefits from the generalized detection model provides deep rich information feature space and ensemble learning [53]. The deep feature maps are generated, the 2$^{nd}$ last fully connected (FC) layer of developed Boosted-BR-STM and customized CNNs and fed to the ML classifier. Ultimately, integrating deep boosting and ensemble classifier improves the DBEL framework generalization ability. In equation (10), $f_{Ensemble}(.)$ an ensemble of ML classifiers makes the final decision from the boosted feature-maps.

$$y_{MLP} = f_{MLP}(\mathbf{x}_{DBF}) \tag{7}$$

$$y_{SVM} = f_{SVM}(\mathbf{x}_{DBF}) \tag{8}$$

$$y_{Adaboost} = f_{Adaboost}(\mathbf{x}_{DBF}) \tag{9}$$

$$y_{Final} = f_{Ensemble}\left(f_{MLP}(\mathbf{x}_{DBF}), f_{SVM}(\mathbf{x}_{DBF}), f_{Adaboost}(\mathbf{x}_{DBF})\right) \tag{10}$$

**Significance of Hybrid Learning**

The training of the proposed deep Boosted-BR-STM and customized CNNs may sometimes cause overfitting. Therefore, the proposed hybrid framework learns the effective discrimination features and improves generalization. Additionally, SVM [48], MLP [51], and AdaBoostM1 [52] represent three distinct ML classifiers that minimize the structural risk. Deep CNNs cause empirical risk by reducing training error using the optimal hyper-parameter selection [48]. The proposed framework significantly reduces both training and test error and thus improves the generalization. Moreover, ensemble learning aims to increase performance and promotes integrating numerous feature spaces into a single rich information feature vector [54].

### 3.4. Utilization of Customized CNNs

Several current CNNs like VGG, ResNet, GoogleNet, DenseNet, and Inception have been adapted to classify parasite malaria images for comparative analysis [23,25,32,35,37,38]. The existing CNN's abstract and final classification layers are customized using additional layers according to the input and target-class dimension of the dataset. CNN models were primarily backpropagation strategies learned from scratch where initial weights were randomly selected. The convolutional layers' initial weights were borrowed from pre-train models using TL to improve the model convergence. In this regard, we used TL to adopt effective model parameters derived from the modified prior CNNs designed to precisely capture the target-domain specific parasite characteristics mostly on the malaria dataset employing improved filter weights acquired from ImageNet [55].

### 4. Experimental Configuration

### 4.1. Dataset

The NIH dataset separated parasitic cells from a thin blood smear slide for research on monitoring and diagnosis [25,56]. The dataset includes samples of falciparum patients collected from the Mahidol Oxford Tropical Medicine Center and Bangladesh's Chittagong Medical College [57]. The institutes mentioned above are proficient slide readers who differentiate between parasitic images and healthy individuals. Plasmodium is present mainly in positive (parasitic) specimens, while the artifact effect is seen in negative (non-parasitic or healthy) samples with staining and contaminants. The malaria dataset distribution for experimental setup is shown in Table 2 and Figure 4, pictorially depicting the parasite and non-parasitic or healthy RBC samples.

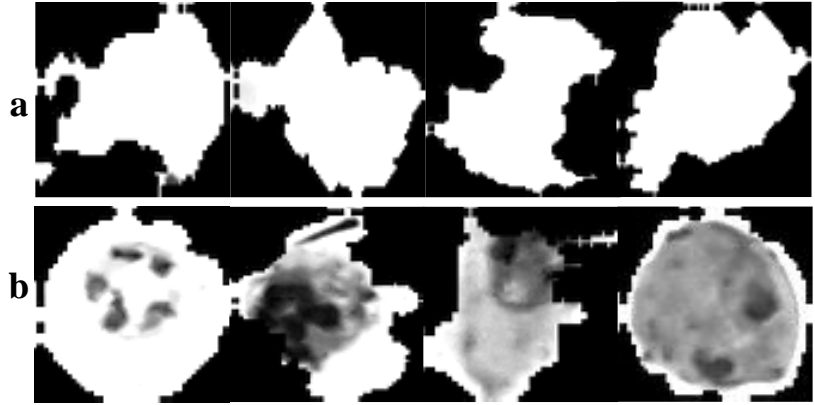

**Figure 4** Panel (a, b) DWT enhanced parasitic and normal samples, respectively.

### 4.2. Model Training

The training and testing parts of the dataset are divided into a 70:30% ratio. Additionally, cross-validation is executed in the training phase, which is partitioned 80:20 into training and validation sets during model training. The training set includes the data-validation set when optimum parameters are picked via hold-out cross-validation. Table 3 lists the selected optimal parameters in detail. The MATLAB-2022a tool was used to create the modified CNNs. The tests were carried out on an NVIDIA-GeForce GTX-T-Dell PC with 32-GB RAM and CUDA support. CNNs required nearly 12–24 hours during training or 1–2 hours for each epoch.

### 4.3. Performance Evaluation

Standard performance criteria are used to evaluate the efficiency of the detection CNNs and the developed hybrid learning framework. The detection measures, including Accuracy, Precision, Sensitivity, Specificity, and F-score, are used as optimization metrics for evaluating the technique's performance. These measurements, along with a mathematical explanation and abbreviation, are explained in equations 11 to 15.

$$\text{Accuracy} = \frac{\text{Classified Correctly}}{\text{Total Samples}} \times 100 \qquad (11)$$

$$\text{Precison} = \frac{\text{Classified Parasitic}}{\text{Classified Parasitic} + \text{Incorrectly Classified Parasitic}} \qquad (12)$$

$$\text{Sensitivity} = \frac{\text{Classified Parasitic}}{\text{Total Parasitic Samples}} \qquad (13)$$

$$\text{Specificity} = \frac{\text{Classified Healty}}{\text{Total Healty Individuals}} \tag{14}$$

$$F - \text{Score} = 2\frac{(\text{Sen x Pre})}{\text{Sen+Pre}} \tag{15}$$

**Table 2** Benchmarked NIH Malaria Detail.

| Characteristics | Overview |
|---|---|
| Total | 27.5k Samples |
| Non-Parasite (Healthy/Impurities) | 13.8k Samples |
| Parasite | 13.8k Samples |
| Train, Validation (70%) | (9.6k, 9.6k) |
| Test (30%) | (4.1k, 4.1k) |
| Input Image | 164 x 164 x 3 |
| Improved Image Size | 82 x 82 x 1 |

**Table 3** Hold-out based optimal hyper-parameter selection.

| Hyper-parameter | Values |
|---|---|
| Learning-rate ($\alpha$) | $10^{-3}$ |
| Optimizer | SGD |
| Epoch | 10 |
| Momentum | 0.90 |
| Loss | Cross Entropy |

## 5. Results And Discussion

This study uses the original and DWT-improved datasets to establish a novel detection DBEL framework system for malaria and distinguish parasite malaria sufferers from healthy individuals. Moreover, a novel Boosted-BR-STM is developed for testing the usefulness of boosting and boundary-region based STM to forecast affected parasite cells in RBC-thin-smears microscopy samples, and its effectiveness is contrasted with modified existing CNNs. Furthermore, the existing CNNs are upgraded, deployed in the TL, and trained from scratch. Table 4 assesses the developed malaria detection employing defined performance metrics.

### 5.1. Performance Analysis of Detection Results

#### 5.1.1. Enhanced Dataset Evaluation

Impurities, staining, and noise artifacts in the original malaria dataset generated a striking likeness between the normal and parasites sample. In this regard, an enhancement process is essential to remove stains/impurities and impulsive noise from blood smear images. In this case, DWT reduces noise effects, stain impurity, and computing time, attaining streamlined, improved feature channels. Moreover, DWT coefficients preserve important low-resolution and diagonal features that aid in differentiating malaria-infected RBCs from healthy samples. Employing CNN techniques on improved DWT image outperformed the standard dataset in terms of Accuracy (1.86-3.1%), F-Score (1.8-2.9%), Sensitivity (0.60-2.2%), Precision (2.40-5.50%), Specificity (2.6-6.1%), as shown in Table 4 and Figure 5. The minimum, average, and maximum performance for each parameter is demonstrated in Figure 5.

Moreover, DWT is frequently employed as an enhancement approach for data cleaning, significant feature-map generation, and improved classification framework performance.

### 5.1.2. The Proposed Boosted-BR-STM

The developed Boosted-BR-STM improved the generalizability of the proposed DBEL scheme compared to early techniques and yielded better results on a DWT-enhanced test dataset. Table 5 and Figure 5 show a significant enhancement in the developed Boosted-BR-STM to forecast plasmodium falciparum-infected patients. The Boosted-BR-STM exploits homogenous and boundary-driven parasite patterns in the STM block, boosting through STM and residual learning, and TL to improve the Sensitivity and F1-score. Region homogenous and boundary characteristics aid in learning the distorted parasitic samples. Moreover, TL-based generated feature-maps and SB can capture subtle contrast and texture variation across artifact and parasite samples.

The data augmentation strategies employed in the training portion yielded performance improvement and model generalization. Moreover, Accuracy and Precision are attained by decreasing false positives, which reduces the medical staff's workload significantly. Tables demonstrate the Boosted-BR-STM fared better than reported approaches using the NIH dataset. The Boosted-BR-STM performs better than other methods; Accuracy values range from (1.16-7.44%), F-score (1.2-7.5%), and Sensitivity (2.5-4.8%) (Figure 5). The developed Boosted-BR-STM lower False Negative (FN=71) and improved Sensitivity relative to the finest, most reputable TL-based DenseNet CNN (FN=131) on unseen DWT enhanced dataset, as demonstrated in Tables 5 and 6.

### 5.1.3. The Proposed Hybrid Learning

The malaria parasite samples are diagnosed using a new deep-boosted feature-map and ensemble learning (DBEL). In this regard, three competing ML classifiers are ensembled with feature vectors of the proposed Boosted-BR-STM that contribute as feature extractors. Moreover, the malaria parasitic images are detected by extracting the deep features from existing CNNs and providing them to ML classifiers in the DHML scheme. The significance of exploiting deep features is identified and compared with Softmax-based evaluation. The DHML using TL-based fine-tuned existing CNNs scheme outperforms Softmax-based evaluation in terms of Accuracy (0.47-1.7%), F-Score (0.50-1.6%), Precision (0.50-3.6%), Specificity (0.5-4%), as illustrated in Table 7.

### 5.1.4. The proposed DBEL Framework

The hybrid learning approach evaluates the performance of deep-boosted feature maps and ensemble learning. The proposed DBEL outperformed the customized CNNs by providing a deep-boosted feature map to the majority voting-based ensembled classifiers. Combining boosted deep feature maps and utilizing three classifiers creates a hybrid of diverse feature spaces and ensemble learning. Boosting learning aims to increase performance and promotes integrating numerous feature spaces into a single rich information feature vector. Moreover, the ensemble classifiers' improves the proposed framework's differentiation capacity by merging deep feature maps to construct the boosted and diverse feature space.

**Table 4** Performance of Boosted-BR-STM and current CNNs on unseen dataset.

| Models | Trained-Scratch (Original-Data) | | | | |
|---|---|---|---|---|---|
| | Accuracy% | F-score | Sensitivity | Precision | Specificity |
| ShuffleNet | 92.85 | 0.931 | 0.965 | 0.899 | 0.892 |
| ResNet-50 | 94.96 | 0.951 | 0.980 | 0.924 | 0.919 |
| SqueezeNet | 95.16 | 0.953 | 0.976 | 0.931 | 0.928 |
| Inceptionv3 | 95.40 | 0.955 | 0.980 | 0.931 | 0.928 |
| ResNet-18 | 95.66 | 0.958 | 0.986 | 0.931 | 0.927 |
| Xception | 95.22 | 0.953 | 0.969 | 0.938 | 0.936 |
| DenseNet-201 | 95.73 | 0.958 | 0.978 | 0.939 | 0.936 |
| VGG-19 | 95.74 | 0.958 | 0.966 | 0.950 | 0.949 |
| VGG-16 | 95.79 | 0.958 | 0.965 | 0.951 | 0.95 |
| GoogleNet | 95.23 | 0.953 | 0.960 | 0.946 | 0.945 |
| **Proposed Boosted-BR-STM** | **96.67** | **0.967** | **0.986** | **0.949** | **0.901** |

**Table 5** Performance comparison of Boosted-BR-STM and current CNNs on unseen dataset.

| Models | Trained-Scratch on Enhanced Data (DWT) | | | | |
|---|---|---|---|---|---|
| | Accuracy% | F-score | Sensitivity | Precision | Specificity |
| ShuffleNet | 94.71 | 0.949 | 0.976 | 0.923 | 0.918 |
| SqueezeNet | 95.34 | 0.954 | 0.969 | 0.940 | 0.938 |
| Inceptionv3 | 95.43 | 0.955 | 0.966 | 0.944 | 0.942 |
| ResNet-50 | 95.55 | 0.956 | 0.969 | 0.944 | 0.942 |
| Xception | 95.67 | 0.957 | 0.980 | 0.936 | 0.933 |
| ResNet-18 | 95.71 | 0.958 | 0.969 | 0.947 | 0.945 |
| VGG-16 | 95.80 | 0.958 | 0.971 | 0.946 | 0.945 |
| DenseNet-201 | 95.84 | 0.959 | 0.973 | 0.945 | 0.944 |
| GoogleNet | 95.85 | 0.959 | 0.982 | 0.938 | 0.935 |
| VGG-19 | 95.95 | 0.960 | 0.966 | 0.954 | 0.953 |
| **Proposed Boosted-BR-STM** | **97.51** | **0.975** | **0.982** | **0.968** | **0.968** |

**Table 6** TL-based existing CNNs performance analysis.

| Models | TL-based CNNs | | | | |
|---|---|---|---|---|---|
| | Accuracy% | F-score | Sensitivity | Precision | Specificity |
| SqueezeNet | 95.08 | 0.952 | 0.978 | 0.927 | 0.923 |
| ShuffleNet | 95.71 | 0.957 | 0.971 | 0.944 | 0.943 |
| VGG-19 | 95.79 | 0.958 | 0.969 | 0.948 | 0.947 |
| Inceptionv3 | 95.91 | 0.959 | 0.971 | 0.948 | 0.947 |
| GoogleNet | 96.03 | 0.962 | 0.983 | 0.941 | 0.938 |
| Xception | 96.07 | 0.961 | 0.971 | 0.951 | 0.95 |
| ResNet-50 | 96.17 | 0.962 | 0.974 | 0.951 | 0.949 |
| VGG-16 | 96.2 | 0.962 | 0.975 | 0.95 | 0.949 |
| ResNet-18 | 96.21 | 0.963 | 0.985 | 0.942 | 0.94 |
| DenseNet-201 | 96.31 | 0.963 | 0.968 | 0.958 | 0.958 |

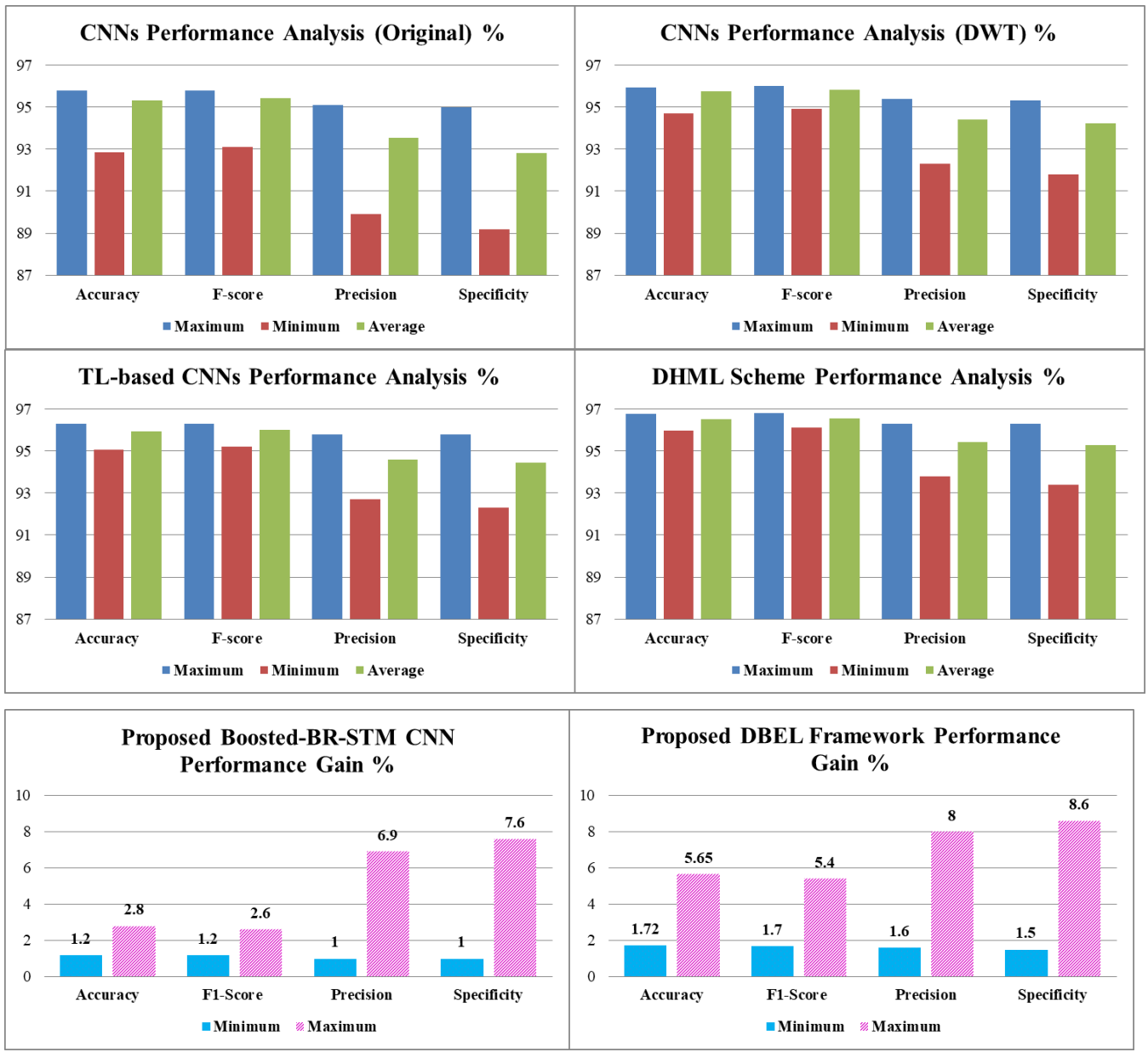

**Figure 5** The proposed Boosted-BR-STM CNN and DBEL framework performance gain over the existing detection. Moreover, an enhanced DWT samples considerably improved performance.

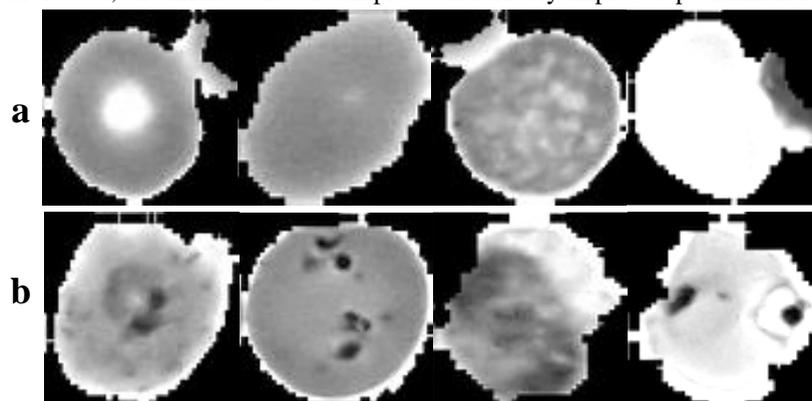

**Figure 6** The enhanced stain artifacts and parasitic samples are shown in panels (a) and (b). The parasite images have a high degree of resemblance to artifacts, resulting in a high rate of miss detection.

**Table 7** Performance of TL-based current CNNs Deep Feature and ML.

| Models | TL-based DHML Scheme | | | | |
|---|---|---|---|---|---|
| | Accuracy% | F-score | Sensitivity | Precision | Specificity |
| SqueezeNet | 95.96 | 0.961 | 0.985 | 0.938 | 0.934 |
| ShuffleNet | 96.26 | 0.963 | 0.969 | 0.957 | 0.956 |
| VGG-19 | 96.34 | 0.964 | 0.974 | 0.954 | 0.953 |
| Inceptionv3 | 96.48 | 0.965 | 0.975 | 0.956 | 0.955 |
| GoogleNet | 96.57 | 0.966 | 0.975 | 0.957 | 0.956 |
| Xception | 96.65 | 0.967 | 0.980 | 0.955 | 0.953 |
| ResNet-50 | 96.65 | 0.967 | 0.983 | 0.952 | 0.950 |
| VGG-16 | 96.69 | 0.967 | 0.971 | 0.963 | 0.963 |
| ResNet-18 | 96.70 | 0.968 | 0.982 | 0.954 | 0.952 |
| DenseNet-201 | 96.78 | 0.968 | 0.978 | 0.958 | 0.957 |
| **Comparative analysis with the existing work on NIH dataset** | | | | | |
| Dense-Net [23] | 90.54 | 0.9050 | 0.9400 | 0.9080 | 0.8770 |
| Inception [23] | 93.06 | 0.9306 | 0.9300 | 0.9306 | 0.9310 |
| Xception [23] | 94.94 | 0.9490 | 0.9260 | 0.9510 | 0.9750 |
| Customized ResNet-50 [38] | 95.40 | --- | --- | --- | --- |
| Seq-CNN [25] | 95.90 | 0.9590 | 0.9470 | --- | 0.9720 |
| VGG-16 [37] | 95.96 | 0.9560 | --- | 0.9680 | --- |
| **Various Malaria Dataset** | | | | | |
| Modified YOLOv3 [40] | 95.46 | --- | --- | --- | --- |
| Modified YOLOv4 [40] | 96.14 | --- | --- | --- | --- |
| YOLOv5 [41] | 95.20 | --- | --- | --- | --- |
| DarkNet-53 [41] | 96.02 | --- | --- | --- | --- |

**Table 8** Performance analysis of the proposed frameworks.

| Framework | The Proposed DBEL Framework Work | | | | |
|---|---|---|---|---|---|
| | Accuracy% | F-score | Sensitivity | Precision | Specificity |
| **Deep Boosted Machine Learning** | | | | | |
| SVM | 97.73 | 0.977 | 0.985 | 0.970 | 0.969 |
| AdaboostM1 | 97.92 | 0.979 | 0.988 | 0.971 | 0.971 |
| MLP | 98.08 | 0.981 | 0.988 | 0.974 | 0.973 |
| **Deep Boosted Ensemble Learning** | | | | | |
| SVM + AdaboostM1 | 98.21 | 0.982 | 0.991 | 0.974 | 0.973 |
| AdaboostM1+MLP | 98.26 | 0.982 | 0.99 | 0.975 | 0.975 |
| MLP+ SVM | 98.34 | 0.983 | 0.991 | 0.976 | 0.976 |
| **Proposed DBEL (SVM + AdaboostM1+ MLP)** | **98.50** | **0.985** | **0.992** | **0.979** | **0.978** |

The individual models within the ensemble like SVM, MLP, AdaBoostM1 are chosen because they have distinct characteristics and strengths. SVM learn optimal decision boundaries, MLP capture complex non-linear relationships, and AdaBoostM1 enhances the classification of challenging instances. The selection of these classifiers is based on their inherent ability to minimize structural risk, and this contributes to the improved performance. The proposed DBEL framework outperformed the existing techniques for classifying malaria parasite samples in terms of Accuracy (1.72-5.65%), F-Score (1.70-5.4%), Precision (1.60-8%), Specificity (1.5-8.6%), as shown in Table 8 and Figure 5. Finally, the proposed DBEL framework further reduced the FN (34), as contrasted to the best and most reputable DenseNet-201. However, a few samples were missed due to a similarity between malaria-infected and healthy people owing to impurity, stained, or noise anomalies in non-parasitic instances, as seen in Figure 6. In the summarized way, we have introduced several innovative concepts to systematically enhance the framework's performance and generalization

**The proposed Deep Boosted-BR-STM CNN novel Concepts**

- Boundary and regional feature extraction and STM Block integration
- Optimal feature map extraction and concatenation of diverse maps to achieve boosted channels through TL-based auxiliary channel generation
- Channel boosting through TL-based auxiliary channel generation

**Deep Hybrid and Ensemble Learning**

- Extracting deep features from the proposed Boosted-BR-STM CNN and utilizing ML Classifiers
- Exploiting proposed deep features and Ensemble of ML Classifiers for the final malaria parasite detection.

### 5.1.5. Customized CNNs

TL is an essential DL technique to perform satisfactorily instead of training CNN from scratch. TL adopted custom CNNs utilizing DWT-enriched data outperformed learning from scratch, as shown in Table 6. The TL-based fine-tuned existing CNNs perform better than those learned from scratch and existing techniques: Accuracy (0.36-1.6%), F1-Score (0.30-1.4%), Sensitivity (0.30-1.9%), Precision (0.40-3.5%), and Specificity (0.5-4%). Tables 5-6 and 8 show the comparative study with customized deep-CNN models, along with Figure 5. TL enhanced the CNN learning convergence through pattern initialization and thus, lowering the miss-classification rate while maintaining precision. This demonstrates TL's ability to analyze medical image fields, especially malaria detection.

### 5.2. Feature Space Visualization

The distinct class samples enhanced the ability of model learning and robustness. Therefore, the principal components analysis (PCA) aims to depict the significant discriminative features. The top-3 (PC1 vs. PC2 and PC1 vs. PC3) distinct components of the test samples are displayed in Figure 8, with distinctive coloring for every feature point for distinguishing parasite-containing samples from non-parasitic ones [58,59]. The feature space is examined to better comprehend the proposed Boosted-STM decision-making BR's behavior in contrast to the existing DenseNet-201 model.

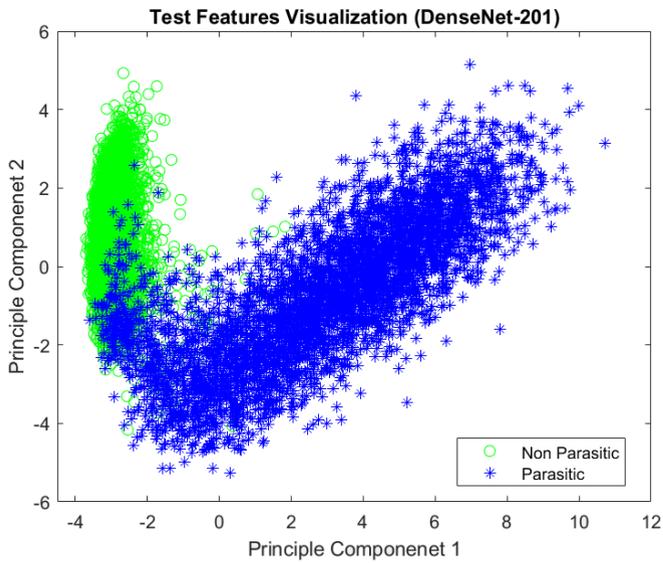
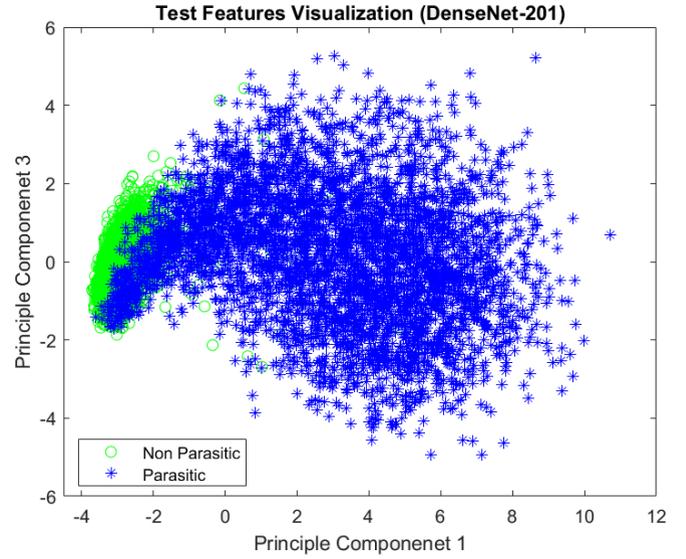
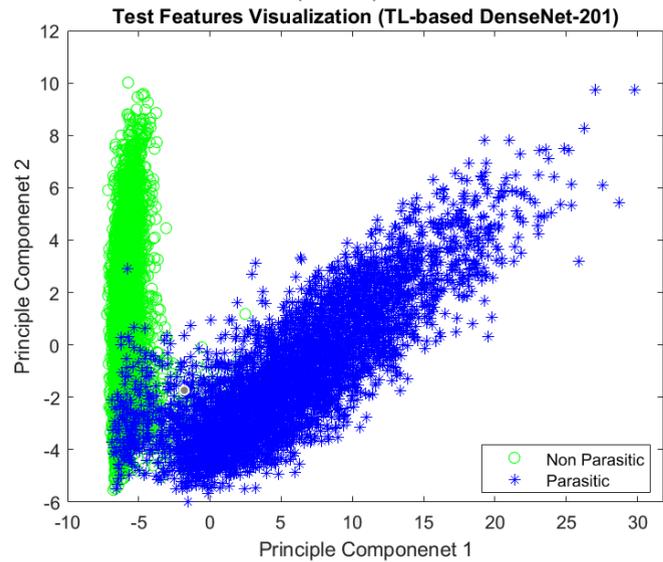
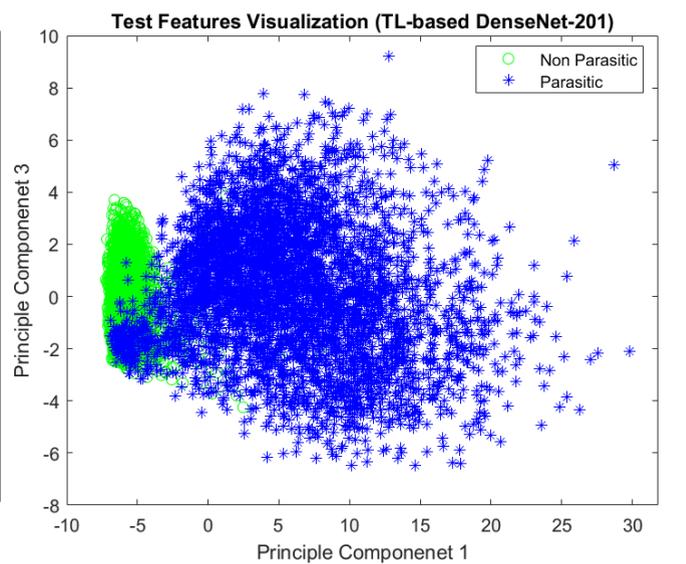
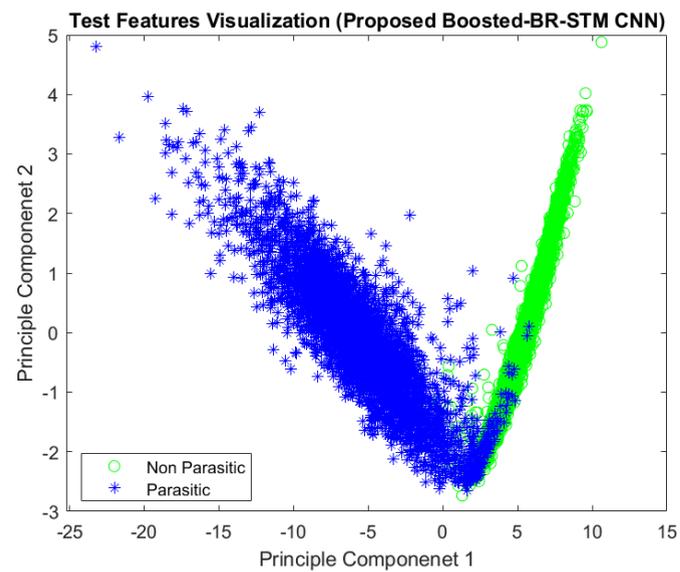
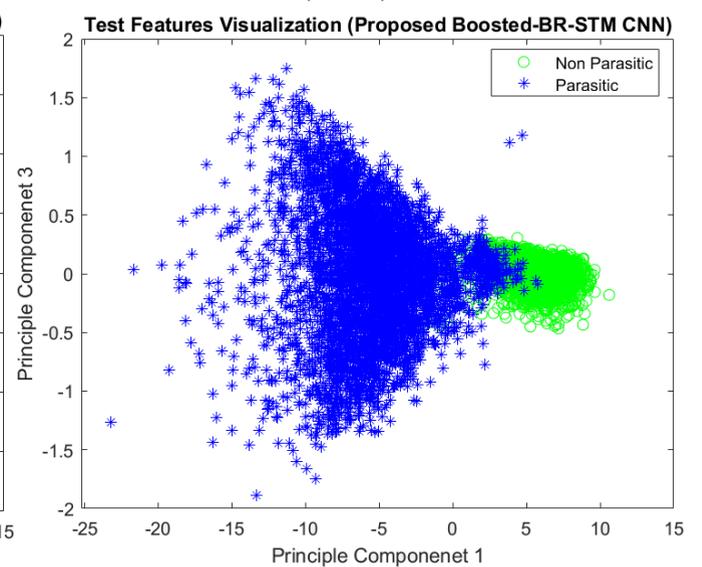

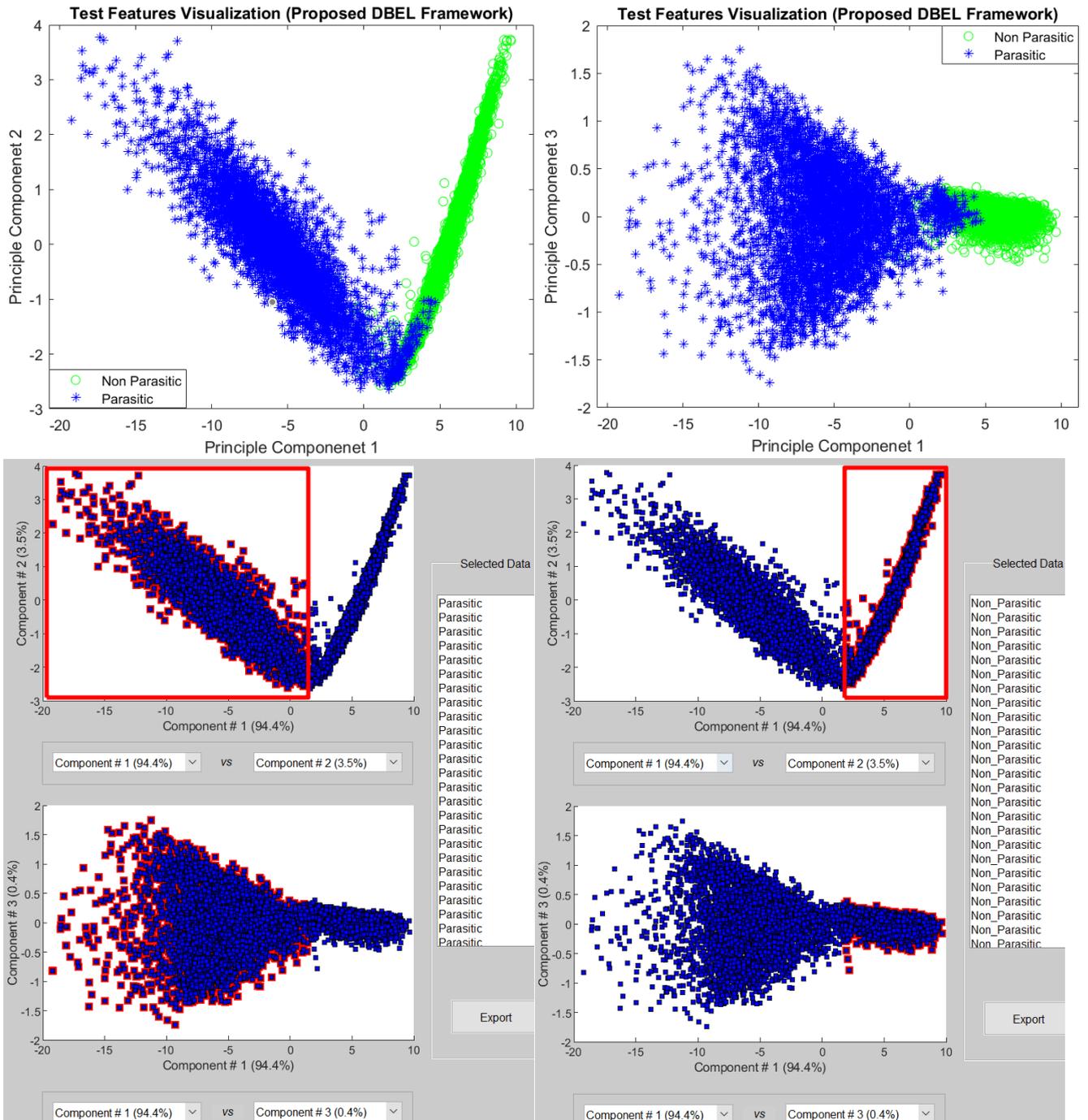

**Figure 7** The visualization for the generated of PC1, PC2, and PC3 distinct features for the proposed DBEL, Boosted-BR-STM, and DenseNet on DWT-enriched datasets.

The developed Boosted-BR-STM and DBEL-generated feature space depict the discrimination among parasite and healthy samples, as illustrated in Figure 7. The Boosted-BR-STM has the highest evaluating effectiveness, according to the 2D PC plots. While the current CNN DenseNet-201 has discovered that feature spaces are cluttered and inaccurately unable to distinguish between two classes using the original data. Additionally, compared to DenseNet-201 trained from scratch using improved DWT data, TL-based DenseNet-201 better indicates the instance class.

## 5.3. Graphical Analysis

The Sensitivity and Precision metrics for assessing the models are critical for malaria parasite detection. Therefore, Precision-Recall (PR) and Receiver-Operating Curve (ROC) are drawn for the developed DBEL framework, Boosted-BR-STM and customized CNNs on test data. These graphs depict the capability of the classifier to differentiate across various potential ranges. The detection cut-off of the best positive class classifier can be considerably accessed via ROC and PR curves; results are shown in Figure 8 [60]. The Boosted-BR-STM reduced false-negative or miss-classified parasitic compared to existing CNNs on an enhanced dataset with a PR AUC=99.20% [61]. Moreover, the developed DBEL framework maintains minimal false-positives by detecting malarial infections with strong Sensitivity due to its high ROC-AUC value.

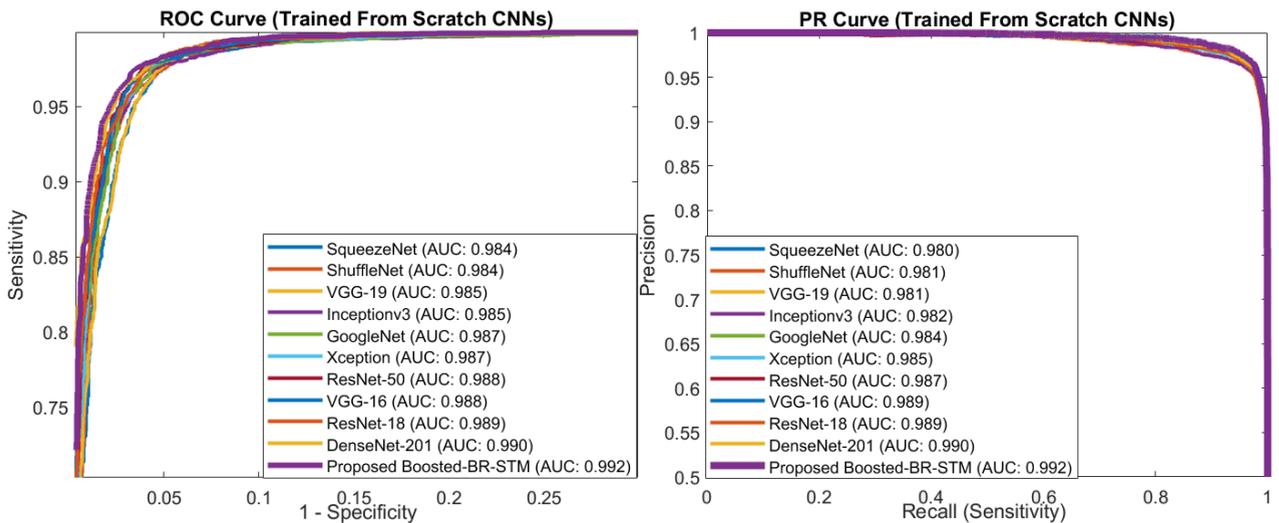

**Figure 8** The developed Boosted-BR-STM with TL-based fine-tuned CNN's ROC and PR curves on original and DWT enriched data.

## 6. Conclusion

Early detection of malaria can be treated properly and quickly enough to prevent irregular disability. The proposed hybrid DBEL framework stack deep boosted features of the developed residual Boosted-BR-STM CNN and ensemble learning to detect RBC thin smears microscope images of malaria patients. The proposed DBEL framework benefits from DWT enhancements, data augmentation, incorporating TL, inherent discriminative deep features from the developed Boosted-BR-STM, and ensemble learning to improve detection performance. Moreover, the developed Boosted-BR-STM employs TL for the diverse rich information feature-maps generation, residual learning, and SB ideas in STM block, enhancing the ability to learn homogeneity, and parasitic structural patterns. Moreover, residual learning systematically extracts features, starting with basic image-level features to more intricate texture-based differences. This novel approach allows for the acquisition of boosted features at varying levels of granularity. The proposed framework achieved an Accuracy of 98.50%, an AUC of 0.996, an F-score of 0.985, and a Sensitivity of 0.992. Furthermore, the proposed Boosted-BR-STM CNN and framework excelled in existing techniques, experimentally in accurately detecting malaria-infected patients. We have assessed our proposed framework's runtime compared to manual analysis, gauging its efficiency and time-saving potential.  It's worth noting that our system's testing time for each parasitic sample averaged 5-7 sec. In the future, the proposed DBEL is intended to help healthcare practitioners by developing application to automatically screen, identify, and stage of parasitic malaria patients for clinical examination, additionally impurity, stained, or noise anomalies in non-parasitic instances. Moreover, the developed framework can be employed in diagnosing monkeypox, brain tumors, lung cancer, and breast cancer using medical images.

**Acknowledgment**

The authors extend their appreciation to the King Salman center For Disability Research for funding this work through Research Group no KSRG-2023-021. We thank Sultan Qaboos University Muscat, Oman, and the Department of Computer Systems Engineering, University of Engineering and Applied Sciences (UEAS), Swat, Pakistan for providing the necessary facilities in carrying out this research work.

**Conflicts of interest**: The authors declare that they have no known competing financial interests or personal relationships that could have appeared to influence the work reported in this paper.

**Informed Consent Statement**
Not applicable.

**Data Availability Statement**

The standard benchmark NIH Malaria dataset has been collected from the standard medical centers and made available in the standard open-access Kaggle and other repositories.

https://www.kaggle.com/datasets/iarunava/cell-images-for-detecting-malaria

# References

[1] Keleta Y, Ramelow J, Cui L, Li J. Molecular interactions between parasite and mosquito during midgut invasion as targets to block malaria transmission. Npj Vaccines 2021;6. https://doi.org/10.1038/s41541-021-00401-9.

[2] Gupta S, Gazendam N, Farina JM, Saldarriaga C, Mendoza I, López-Santi R, et al. Malaria and the Heart: JACC State-of-the-Art Review. J Am Coll Cardiol 2021;77:1110–21. https://doi.org/10.1016/j.jacc.2020.12.042.

[3] Yimam Y, Nateghpour M, Mohebali M, Afshar MJA. A systematic review and meta-analysis of asymptomatic malaria infection in pregnant women in Sub-Saharan Africa: A challenge for malaria elimination efforts. PLoS One 2021;16. https://doi.org/10.1371/journal.pone.0248245.

[4] Tegegne Y, Worede A, Derso A, Ambachew S. The Prevalence of Malaria among Children in Ethiopia: A Systematic Review and Meta-Analysis. J Parasitol Res 2021;2021. https://doi.org/10.1155/2021/6697294.

[5] World Health Organization (WHO). WHO Malaria Policy Advisory Group ( MPAG ) meeting 2021:13–4.

[6] Abbas N, Saba T, Rehman A, Mehmood Z, Javaid N, Tahir M, et al. Plasmodium species aware based quantification of malaria parasitemia in light microscopy thin blood smear. Microsc Res Tech 2019;82:1198–214. https://doi.org/10.1002/jemt.23269.

[7] Yoon J, Jang WS, Nam J, Mihn DC, Lim CS. An automated microscopic Malaria parasite detection system using digital image analysis. Diagnostics 2021;11. https://doi.org/10.3390/diagnostics11030527.

[8] Barber BE, William T, Grigg MJ, Yeo TW, Anstey NM. Limitations of microscopy to differentiate Plasmodium species in a region co-endemic for Plasmodium falciparum, Plasmodium vivax and Plasmodium knowlesi. Malar J 2013;12:8. https://doi.org/10.1186/1475-2875-12-8.

[9] Mukry SN, Saud M, Sufaida G, Shaikh K, Naz A, Shamsi TS. Laboratory diagnosis of malaria: Comparison of manual and automated diagnostic tests. Can J Infect Dis Med Microbiol 2017;2017. https://doi.org/10.1155/2017/9286392.

[10] Maity M, Gantait K, Mukherjee A, Chatterjee J. Visible spectrum-based classification of malaria blood samples on handheld spectrometer. I2MTC 2019 - 2019 IEEE Int Instrum Meas Technol Conf Proc 2019;2019-May. https://doi.org/10.1109/I2MTC.2019.8826860.

[11] Microwave A, Engineering E. Software / Diagnostic Manual n.d.:1–8.

[12] Somasekar J, Sharma A, Madhusudhana Reddy N, Padmanabha Reddy YCA. Image analysis for automatic enumeration of rbc infected with plasmodium parasites-implications for malaria diagnosis. Adv Math Sci J 2020;9:1229–37. https://doi.org/10.37418/amsj.9.3.48.

[13] Molina A, Rodellar J, Boldú L, Acevedo A, Alférez S, Merino A. Automatic identification of malaria and other red blood cell inclusions using convolutional neural networks. Comput Biol Med 2021;136:104680. https://doi.org/10.1016/j.compbiomed.2021.104680.

[14] Leckenby J, Li H, Negus K, Pickering M, Adorno T, Horkheimer M, et al. A semi-automatic method for quantification and classification of erythrocytes infected with malaria parasites in microscopic images. J Biomed Inform 2009;42:296–307. https://doi.org/10.1016/j.jbi.2008.11.005.

[15] Krishnadas P, Sampathila N. Automated Detection of Malaria implemented by Deep Learning in Pytorch. 2021 IEEE Int. Conf. Electron. Comput. Commun. Technol., IEEE; 2021, p. 01–5. https://doi.org/10.1109/CONECCT52877.2021.9622608.

[16] Kalkan SC, Sahingoz OK. Deep learning based classification of malaria from slide images. 2019 Sci Meet Electr Biomed Eng Comput Sci EBBT 2019 2019. https://doi.org/10.1109/EBBT.2019.8741702.

[17] Baroni L, Salles R, Salles S, Guedes G, Porto F, Bezerra E, et al. An analysis of malaria in the

Brazilian Legal Amazon using divergent association rules. J Biomed Inform 2020;108:103512. https://doi.org/10.1016/j.jbi.2020.103512.

[18] Asam M, Khan SH, Akbar A, Bibi S, Jamal T, Khan A, et al. IoT malware detection architecture using a novel channel boosted and squeezed CNN. Sci Rep 2022;12:15498. https://doi.org/10.1038/s41598-022-18936-9.

[19] Zahoora U, Khan A, Rajarajan M, Khan SH, Asam M, Jamal T. Ransomware detection using deep learning based unsupervised feature extraction and a cost sensitive Pareto Ensemble classifier. Sci Rep 2022;12:15647. https://doi.org/10.1038/s41598-022-19443-7.

[20] Khan A, Khan SH, Saif M, Batool A, Sohail A, Khan MW. A Survey of Deep Learning Techniques for the Analysis of COVID-19 and their usability for Detecting Omicron 2022.

[21] Du X, Wang X, Xu F, Zhang J, Huo Y, Ni G, et al. Morphological components detection for super-depth-of-field bio-micrograph based on deep learning. Microscopy 2022;71:50–9. https://doi.org/10.1093/jmicro/dfab033.

[22] Zafar MM, Rauf Z, Sohail A, Khan AR, Obaidullah M, Khan SH, et al. Detection of tumour infiltrating lymphocytes in CD3 and CD8 stained histopathological images using a two-phase deep CNN. Photodiagnosis Photodyn Ther 2022;37:102676. https://doi.org/10.1016/j.pdpdt.2021.102676.

[23] Maqsood A, Farid MS, Khan MH, Grzegorzek M. Deep malaria parasite detection in thin blood smear microscopic images. Appl Sci 2021;11:1–19. https://doi.org/10.3390/app11052284.

[24] Lin M, Huang C, Chen R, Fujita H, Wang X. Directional correlation coefficient measures for Pythagorean fuzzy sets: their applications to medical diagnosis and cluster analysis. Complex Intell Syst 2021;7:1025–43. https://doi.org/10.1007/s40747-020-00261-1.

[25] Rajaraman S, Antani SK, Poostchi M, Silamut K, Hossain MA, Maude RJ, et al. Pre-trained convolutional neural networks as feature extractors toward improved malaria parasite detection in thin blood smear images. PeerJ 2018;2018. https://doi.org/10.7717/peerj.4568.

[26] Das DK, Ghosh M, Pal M, Maiti AK, Chakraborty C. Machine learning approach for automated screening of malaria parasite using light microscopic images. Micron 2013;45:97–106. https://doi.org/10.1016/j.micron.2012.11.002.

[27] Sarkar RP, Maiti A. Investigation of dataset from diabetic retinopathy through discernibility-based k-NN algorithm. Adv Intell Syst Comput 2019;812:93–100. https://doi.org/10.1007/978-981-13-1540-4_10.

[28] T. Colwell WBMCS. Automated Detection of P. falciparum Using Machine Learning Algorithms with Quantitative Phase Images of Unstained Cells. PLoS One 2016;11:e0163045. https://doi.org/10.1371/journal.pone.0163045.

[29] T. G, J.H. K, H. B, S.J. L. Machine learning-based in-line holographic sensing of unstained malaria-infected red blood cells. J Biophotonics 2018;11:e201800101.

[30] Mehanian C, Jaiswal M, Delahunt C, Thompson C, Horning M, Hu L, et al. Computer-Automated Malaria Diagnosis and Quantitation Using Convolutional Neural Networks. Proc - 2017 IEEE Int Conf Comput Vis Work ICCVW 2017 2017;2018-Janua:116–25. https://doi.org/10.1109/ICCVW.2017.22.

[31] Bibin D, Nair MS, Punitha P. Malaria Parasite Detection from Peripheral Blood Smear Images Using Deep Belief Networks. IEEE Access 2017;5:9099–108. https://doi.org/10.1109/ACCESS.2017.2705642.

[32] Var E, Boray Tek F. Malaria Parasite Detection with Deep Transfer Learning. UBMK 2018 - 3rd Int Conf Comput Sci Eng 2018:298–302. https://doi.org/10.1109/UBMK.2018.8566549.

[33] Dong Y, Jiang Z, Shen H, David Pan W, Williams LA, Reddy VVB, et al. Evaluations of deep convolutional neural networks for automatic identification of malaria infected cells. 2017 IEEE EMBS Int Conf Biomed Heal Informatics, BHI 2017 2017:101–4.

https://doi.org/10.1109/BHI.2017.7897215.
[34]  Lenet-5, convolutional neural networks 2015.
[35]  Szegedy C, Wei Liu, Yangqing Jia, Sermanet P, Reed S, Anguelov D, et al. Going deeper with convolutions. 2015 IEEE Conf. Comput. Vis. Pattern Recognit., vol. 07-12- June, IEEE; 2015, p. 1–9. https://doi.org/10.1109/CVPR.2015.7298594.
[36]  Hung, Jane, Allen Goodman, Stefanie Lopes, Carpenter A. Applying Faster R-CNN for Object Detection on Malaria Images. J R Stat Soc Ser A Stat Soc 2013;175:417–33.
[37]  Huq A, Pervin MT. Robust Deep Neural Network Model for Identification of Malaria Parasites in Cell Images. 2020 IEEE Reg 10 Symp TENSYMP 2020 2020:1456–9. https://doi.org/10.1109/TENSYMP50017.2020.9230832.
[38]  Reddy ASB, Juliet DS. Transfer Learning with ResNet-50 for Malaria Cell-Image Classification. 2019 Int. Conf. Commun. Signal Process., Boston, MA: IEEE; 2019, p. 0945–9. https://doi.org/10.1109/ICCSP.2019.8697909.
[39]  Maity M, Jaiswal A, Gantait K, Chatterjee J, Mukherjee A. Quantification of malaria parasitaemia using trainable semantic segmentation and capsnet. Pattern Recognit Lett 2020;138:88–94. https://doi.org/10.1016/j.patrec.2020.07.002.
[40]  Abdurahman F, Fante KA, Aliy M. Malaria parasite detection in thick blood smear microscopic images using modified YOLOV3 and YOLOV4 models. BMC Bioinformatics 2021;22:112. https://doi.org/10.1186/s12859-021-04036-4.
[41]  Zedda L, Loddo A, Di Ruberto C. A Deep Learning Based Framework for Malaria Diagnosis on High Variation Data Set. Ann. Tour. Res., vol. 3, 2022, p. 358–70. https://doi.org/10.1007/978-3-031-06430-2_30.
[42]  Houwen B. Blood film preparation and staining procedures. Clin Lab Med 2002;22:1–14. https://doi.org/10.1016/S0272-2712(03)00064-7.
[43]  Sakthidasan alias Sankaran K, Nagarajan V. Noise Removal Through the Exploration of Subjective and Apparent Denoised Patches Using Discrete Wavelet Transform. IETE J Res 2021;67:843–52. https://doi.org/10.1080/03772063.2019.1569483.
[44]  Pandit P, Anand A. Diagnosis of Malaria Using Wavelet Coefficients and Dynamic Time Warping. Int J Appl Comput Math 2019;5. https://doi.org/10.1007/s40819-019-0614-2.
[45]  Shorten C, Khoshgoftaar TM. A survey on Image Data Augmentation for Deep Learning. J Big Data 2019;6. https://doi.org/10.1186/s40537-019-0197-0.
[46]  Khan SH, Shah NS, Nuzhat R, Majid A, Alquhayz H, Khan A. Malaria parasite classification framework using a novel channel squeezed and boosted CNN. Microscopy 2022. https://doi.org/10.1093/jmicro/dfac027.
[47]  Khan SH, Sohail A, Zafar MM, Khan A. Coronavirus disease analysis using chest X-ray images and a novel deep convolutional neural network. Photodiagnosis Photodyn Ther 2021;35:102473. https://doi.org/10.1016/j.pdpdt.2021.102473.
[48]  Khan SH, Sohail A, Khan A, Hassan M, Lee YS, Alam J, et al. COVID-19 detection in chest X-ray images using deep boosted hybrid learning. Comput Biol Med 2021;137. https://doi.org/10.1016/j.compbiomed.2021.104816.
[49]  Aziz A, Sohail A, Fahad L, Burhan M, Wahab N, Khan A. Channel Boosted Convolutional Neural Network for Classification of Mitotic Nuclei using Histopathological Images. Proc. 2020 17th Int. Bhurban Conf. Appl. Sci. Technol. IBCAST 2020, 2020. https://doi.org/10.1109/IBCAST47879.2020.9044583.
[50]  Khan SH. COVID-19 Detection and Analysis From Lung CT Images using Novel Channel Boosted CNNs 2022. https://doi.org/2209.10963.
[51]  Gardner M., Dorling S. Artificial neural networks (the multilayer perceptron)—a review of applications in the atmospheric sciences. Atmos Environ 1998;32:2627–36.

https://doi.org/10.1016/S1352-2310(97)00447-0.
[52] CAO Y, MIAO Q-G, LIU J-C, GAO L. Advance and Prospects of AdaBoost Algorithm. Acta Autom Sin 2013;39:745–58. https://doi.org/10.1016/S1874-1029(13)60052-X.
[53] Zahoor MM, Qureshi SA, Bibi S, Khan SH, Khan A, Ghafoor U, et al. A New Deep Hybrid Boosted and Ensemble Learning-Based Brain Tumor Analysis Using MRI. Sensors 2022;22:2726. https://doi.org/10.3390/s22072726.
[54] Ganaie MA, Hu M, Tanveer* M, Suganthan* PN. Ensemble deep learning: A review 2021.
[55] Khan SH, Khan A, Lee YS, Hassan M, Jeong WK. Segmentation of shoulder muscle MRI using a new Region and Edge based Deep Auto-Encoder. Multimed Tools Appl 2022. https://doi.org/10.1007/s11042-022-14061-x.
[56] ARUNAVA. Malaria Cell Images Dataset | Kaggle n.d. https://www.kaggle.com/datasets/iarunava/cell-images-for-detecting-malaria (accessed December 20, 2022).
[57] Maude RJ, Hasan MU, Hossain MA, Sayeed AA, Kanti Paul S, Rahman W, et al. Temporal trends in severe malaria in Chittagong, Bangladesh. Malar J 2012;11. https://doi.org/10.1186/1475-2875-11-323.
[58] Lobo SA, Siswadi, Bakhtiar T. Visualization of classified data with kernel principal component analysis. Glob J Pure Appl Math 2015;11:2347–56. https://doi.org/10.31227/osf.io/cbfxu.
[59] Barshan E, Ghodsi A, Azimifar Z, Zolghadri Jahromi M. Supervised principal component analysis: Visualization, classification and regression on subspaces and submanifolds. Pattern Recognit 2011;44:1357–71. https://doi.org/10.1016/j.patcog.2010.12.015.
[60] Hajian-Tilaki K. Receiver operating characteristic (ROC) curve analysis for medical diagnostic test evaluation. Casp J Intern Med 2013;4:627–35.
[61] Boyd K, Eng KH, Page CD. Area under the precision-recall curve: Point estimates and confidence intervals. Lect Notes Comput Sci (Including Subser Lect Notes Artif Intell Lect Notes Bioinformatics) 2013;8190 LNAI:451–66. https://doi.org/10.1007/978-3-642-40994-3_29.